\documentclass[aps,showpacs, prl,twocolumn,superscriptaddress]{revtex4}%
\usepackage{epsfig,dsfont,amssymb,amsmath,amsthm,amsfonts,amsbsy,mathrsfs}
\usepackage{graphicx}
\usepackage{amsmath}
\usepackage{amssymb}
\begin{document}
\title{Systematic enumeration of crystalline networks with only sp$^{2}$ configuration in cubic lattices}
\author{Chaoyu He}
\affiliation{Hunan Key Laboratory for Micro-Nano Energy Materials
and Devices, Xiangtan University, Hunan 411105, P. R. China;}
\affiliation{Laboratory for Quantum Engineering and Micro-Nano
Energy Technology, Faculty of Materials and Optoelectronic Physics,
Xiangtan University, Hunan 411105, P. R. China.}
\author{L. Z. Sun}
\affiliation{Hunan Key Laboratory for Micro-Nano Energy Materials
and Devices, Xiangtan University, Hunan 411105, P. R. China;}
\affiliation{Laboratory for Quantum Engineering and Micro-Nano
Energy Technology, Faculty of Materials and Optoelectronic Physics,
Xiangtan University, Hunan 411105, P. R. China.}
\author{C. X. Zhang}
\affiliation{Hunan Key Laboratory for Micro-Nano Energy Materials
and Devices, Xiangtan University, Hunan 411105, P. R. China;}
\affiliation{Laboratory for Quantum Engineering and Micro-Nano
Energy Technology, Faculty of Materials and Optoelectronic Physics,
Xiangtan University, Hunan 411105, P. R. China.}
\author{J. X. Zhong}
\email{zhong.xtu@gmail.com} \affiliation{Hunan Key Laboratory for
Micro-Nano Energy Materials and Devices, Xiangtan University, Hunan
411105, P. R. China;} \affiliation{Laboratory for Quantum
Engineering and Micro-Nano Energy Technology, Faculty of Materials
and Optoelectronic Physics, Xiangtan University, Hunan 411105, P. R.
China.}
\date{\today}
\pacs{61.50.Ks, 61.66.Bi, 62.50. -p, 63.20. D-}

\begin{abstract}
Systematic enumeration of crystalline networks with some special
topological characters is of considerable interest in both
mathematics and crystallography. Based on the restriction of lattice
in cubic and inequivalent nodes not exceeding three, a simple method
is proposed for systematic searching for three-dimensional
crystalline networks with only sp$^2$-configuration nodes
(C-sp$^2$-TDTCNs). We systematically scan the cubic space groups
from No.195 to No.230 and find many C-sp$^2$-TDTCNs including all
the previously proposed cubic ones. These C-sp$^2$-TDTCNs are
topologically intriguing and can be considered as good templates for
searching carbon crystals with novel properties, predicting high
pressure phases of element nitrogen and designing three-dimensional
hydrocarbon crystals. Structure optimizations are considered by
regrading these C-sp$^2$-TDTCNs as carbon crystals and the
corresponding energetic stability of these carbon crystals are
evaluated, using the the density functional theory (DFT) based
first-principle calculations. Our results are of wide interests in
mathematics, condensed physics, crystallography and material
science.\end{abstract} \maketitle
\section{INTRODUCTION}
\indent Systematic enumeration of crystalline networks with some
special topological characters is of considerable interest in both
mathematics and crystallography. With particular importance, the
three-dimensional (TD) four- and/or three-connected networks
relevant to a wide range of systems (especially to the crystal
structures, such as zeolitic and the well-known diamond) have been
systematically investigated \cite{1, 2, 3, 4, 5, 6, 7, 8, 9, 10,
11}. TD three-connected networks (TDTCNs) have also been
considered\cite{2, 4, 6, 8} as topologically intriguing structures
in mathematics and are also significant in crystallography. The
crystalline ground state of carbon, namely graphite, contains carbon
atoms with perfect sp$^2$-configuration (the ground state of the
binary compound of boron and nitrogen, hexagonal boron nitride, also
prefers such a sp$^2$-configuration). But the graphite structure is
not an exact TD network. Many efforts have been paid on predicting
carbon crystals with sp$^2$-configuration in exact TD networks with
the belief that graphite is not the only way to fill the TD space
with only trigonal sp$^2$-like atoms\cite{bct4, H6, cnt1, s1, s2,
s3, s4, s5, s6, s7, s8, s9, p1, p2, cf, c201, c202, R6, cpl1, cpl2,
fcc64, cnt2, cnt3, k41, k42, k43, cnt4, cnt5}. There are many TDTCNs
\cite{2, 4, 6, 8} satisfying such a requirement and some of them
have been investigated as carbon crystals\cite{bct4, H6, cnt1, s1,
s2, s3, s4, s5, s6, s7, s8, s9, p1, p2, cf, c201, c202, R6, cpl1,
cpl2, fcc64, cnt2, cnt3, k41, k42, k43, cnt4, cnt5} showing novel
physical properties. The second fact implying the signification of
TDTCNs (especially of those with sp$^2$-configuration,
sp$^2$-TDTCNs) is about nitrogen. At ambient condition, element
nitrogen forms the only gas phase of molecular N$_2$ with
nitrogen-nitrogen triple bond. With the progress in high pressure
technology, scientists realized a phase transition from molecular
nitrogen (N$_2$) to an incompressible single-bonded cubic gauche
phase (cg-N)\cite{cgne, cgnt}. In such a novel solid phase, nitrogen
atoms form nontraditional sp$^3$-hybridization where three electrons
bond to three neighbours (fill three sp$^3$-hybridized orbits)
forming a distorted sp$^2$-configuration and the lone pair electrons
fill the fourth sp$^3$-hybridized orbit. Many followed theoretical
predictions\cite{n1, n2, n3, n4, n5, n6} also show that high
pressure nitrogen prefers the TDTCNs with distorted
sp$^2$-configurations. Thirdly, sp$^2$-TDTCNs are also good
templates for designing TD hydrocarbon crystals. Many efforts have
been paid in this direction\cite {ch1, ch2, ch3, ch4, ch5, ch6}. For
example, K4-carbon\cite{k41, k42, k43} to K4-hydrocarbon\cite{ch6},
graphene\cite{graphene} to graphane\cite{ch1} and graphite to
graphane crystals\cite{ch2}.\\
\indent Predictions of the TDTCNs can be found in many previous
works\cite{2, 4, 6, 8} and the predictions of non-graphite TD
sp$^2$-hybridized carbon crystals, according to our knowledge, were
started in 1983 by Hoffmann (bct4-carbon) \cite{bct4}. Since 1983,
many scientists have reported their predictions on TD
sp$^2$-hybridized carbon crystals, such as H6-carbon in
1990\cite{H6}, 6.8$^2$D (polybenzene) and 6.8$^2$P in 1992\cite{p1},
6.8$^2$G and P56 in 1993\cite{p2}, carbon forms in 1995\cite{cf},
C20-carbon in 1997 \cite{c201} and later in 1998 \cite{c202},
FCC-(C$_{28}$)$_2$, FCC-(C$_{36}$)$_2$ and FCC-(C$_{40}$)$_2$ in
1998\cite{s9}, R6-carbon and BCT8-carbon in 1998\cite{R6},
FCC-(C$_{64}$)$_2$ in 2003\cite{fcc64}, K4-carbon in 2008\cite{k41}
and 2009\cite{k42, k43}, C$_{152}$ and C$_{200}$ in 2010\cite{s7},
TD tubular carbons\cite{cnt1, cnt2, cnt3, cnt4, cnt5} and Schwarzite
carbons (P, D, G surface)\cite{p2, s1, s2, s3, s4, s5, s6, s7}. By
combining graph theory with quantum mechanics, Winkler et al.
systematically predicted\cite{cpl1, cpl2} 14 TD sp$^2$-hybridized
carbon crystals with up to six atoms per unit cell. Their results
include previously proposed bct4-carbon\cite{bct4},
H6-carbon\cite{H6}, R6-carbon\cite{R6} and K4-carbon\cite{k41, k42,
k43}. Very recently, with the idea of substitution mentioned by
Sheng et al (T-carbon)\cite{T-carbon}, we have also successfully
constructed\cite{pccp} a simple TD sp$^2$-hybridized carbon crystal
(sp$^2$-diamond) with intriguing configuration, which belongs to the
same space group of diamond and has only one inequivalent atomic
position. However, the number of the TDTCNs is infinite. We can not
enumerate all of them without any restrictions. In this paper, we
report a partial solution to this problem, based on a restriction of
lattice in cubic and inequivalent atomic positions not exceeding
three. With such a restriction and the group theory, a systematic
(not a global) enumeration of cubic TDTCNs with sp$^2$-like
topological configurations (C-sp$^2$-TDTCNs) by
designed-computational-program becomes possible. We have
systematically scanned all the cubic space groups from No.195 to
No.230 and found many C-sp$^2$-TDTCNs including all the previously proposed cubic ones.\\
\section{Method}
Before we introduce the method of our systematic enumeration of
C-sp$^2$-TDTCNs, four simple but very useful rules are discussed.
\textbf{Rule one}, if a C-TDTCN contains only one inequivalent
atomic position, the three neighbours for every atoms are
themselves; \textbf{Rule two}, if a C-TDTCN contains only two
inequivalent atomic positions, its two inequivalent atoms form a
chemical bond. \textbf{Rule three}, if a C-TDTCN contains only three
inequivalent atomic positions, its three inequivalent atoms form an
angle. \textbf{Rule four}, variation of lattice constant for any
cubic crystal does not affect its bond angles. Based on these four
simple rules, we can systematically search C-sp$^2$-TDTCNs with one,
two and three inequivalent atoms by random putting an atom, a bond
and an angle, respectively, in a given cubic lattice in a special
space group with a defined sp$^2$-criterion. Such an idea enhances
the probability of generating C-sp$^2$-TDTCNs in comparison with a
totally random method. A simple procedure implementing such an idea
is shown in Fig1. Predictions of C-sp$^2$-TDTCNs containing more
than three inequivalent atomic positions or belonging to non-cubic
lattice become more complicated (that can also be solved in
principle) and are not considered in our present work.\\
\begin{figure}
  \includegraphics[width=3.5in]{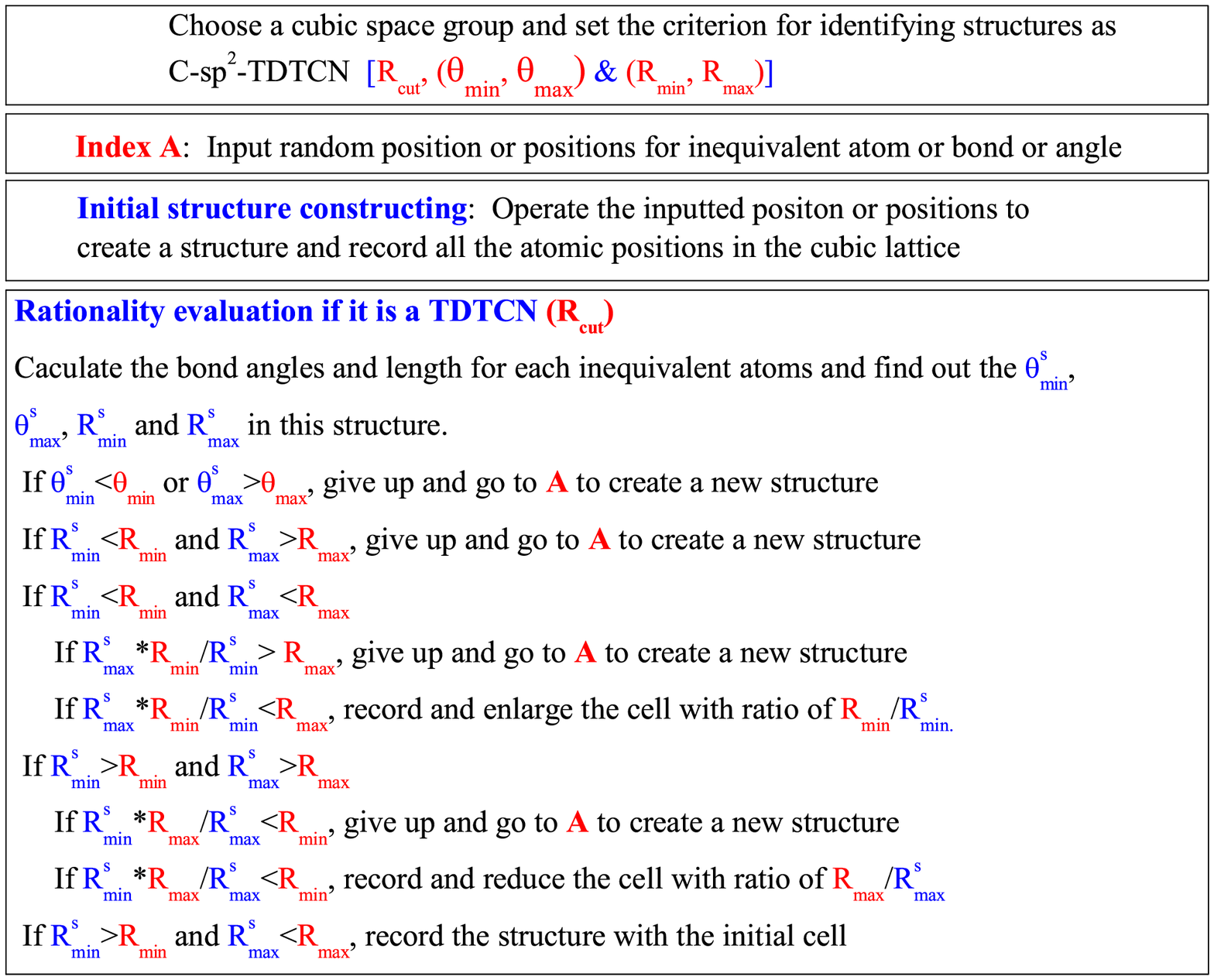}
  \caption{Procedure for systematic searching for C-sp$^2$-TDTCNs}\label{Fig1}
\end{figure}
\indent Based on the procedure shown Fig1, one can systematically
search C-sp$^2$-TDTCNs in the following four simple steps. I),
Choose a cubic lattice with designated lattice constant and space
group and set a sp$^2$-criterion for identifying structures as
C-sp$^2$-TDTCNs, namely the cutoff radius (R$_{cut}$) for the fourth
neighbour, the minima and maxima of bond lengths (R$_{min}$,
R$_{max}$) and bond angles ($\theta$$_{min}$, $\theta$$_{max}$) in
C-sp$^2$-TDTCNs. II), Randomly input the position or positions for
the inequivalent atom or bond or angle in the chosen lattice and
operate them by corresponding symmetry operators to construct a
testing-crystal. III), Calculate the bond lengths and bond angles
for each inequivalent atoms in the created testing-crystal and find
out the minima (R$^s_{min}$, $\theta$$^s_{min}$) and maxima
(R$^s_{max}$, $\theta$$^s_{max}$) for bond lengthes and bond angles,
respectively. IV), Evaluate the rationality of the testing-crystal
according to the following five conditions if it can be considered
as a three-connected network (The requirement of R$_{cut}$ is
satisfied): a) $\theta$$^s_{min}$$<$$\theta$$_{min}$ or
$\theta$$^s_{max}$$>$$\theta$$_{max}$ indicates that such a
structure dose not satisfy the requirements of bond angle. We give
up it directly because further cell enlarging or reducing can not
change the angles according to the rule four mentioned before. b)
R$^s_{min}$$<$R$_{min}$ and R$^s_{max}$$>$R$_{max}$ indicate that
such a structure dose not satisfy both the minimum and maximum
requirements of bond length. We give up it directly because further
cell enlarging (reducing) can not enlarge (reduce) the R$^s_{min}$
(R$^s_{max}$) and reduce (enlarge) the R$^s_{max}$ (R$^s_{min}$),
making the requirements of bond length satisfied. c)
R$^s_{min}$$<$R$_{min}$ and R$^s_{max}$$<$R$_{max}$ indicate that
such a structure dose not satisfy the minimum bond length
requirement. We then enlarge the crystal cell with ratio of
R$_{min}$/R$^s_{min}$ to make the crystal satisfying the minimum
bond length requirement. If the structure is still a three-connected
one and the maximum bond length requirement is satisfied (namely,
R$^s_{max}$*R$_{min}$/R$^s_{min}$$<$R$_{max}$), we record the
enlarged structure (otherwise give it up). d)
R$^s_{min}$$>$R$_{min}$ and R$^s_{max}$$>$R$_{max}$ indicate that
such a structure dose not satisfy the maximum bond length
requirement. We then reduce the crystal cell with ratio of
R$_{max}$/R$^s_{max}$ to make the crystal satisfy the maximum bond
length requirement. If the reduced structure is still
three-connected one and the minimum bond length requirement is still
satisfied (namely, R$^s_{min}$*R$_{max}$/R$^s_{max}$$>$R$_{min}$),
we record the reduced structure (otherwise give it up). e)
R$^s_{min}$$>$R$_{min}$ and R$^s_{max}$$<$R$_{max}$ indicate that
such a structure directly satisfies both maximum and minimum bond
length requirements and we record it directly.\\
\begin{figure}
  \includegraphics[width=3.5in]{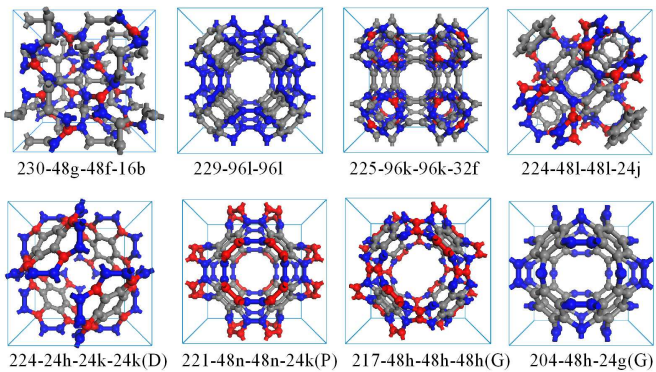}
  \caption{Some selected crystalline views of new C-sp$^2$-TDTCNs found in present work. Different
  colors indicate inequivalent atomic positions and D, G, P indicate the surface types.}\label{Fig1}
\end{figure}
\section{Results and discussion}
\indent In our present work we systematically enumerate the
C-sp$^2$-TDTCNs by the method as discussed above. We name all the
C-sp$^2$-TDTCNs with their space group number and their
corresponding Wyckoff positions. For example, the previously
proposed 6.8$^2$D\cite{p1}, C20-carbon \cite{c201, c202} and P192
\cite{s3}(P8\cite{s5}), containing one, two and three inequivalent
atoms, are named as 224-24i, 225-48h-32f and 229-96l-48k-48k,
respectively. With such a nomenclature, we can know the fundamental
information of a structure, which are useful for identifying if it
has been proposed previously and excluding repeating structures.
Based on such a nomenclature, some C-sp$^2$-TDTCNs possess the same
name due to their commonness in both space group and Wyckoff
positions. We provide their structural characters to distinguish
them. We have systematically scanned all the cubic space groups from
No.195 to No.230 based on the requirements of cutoff radius of 2.0
{\AA}, bond angles distributing in 109$^\circ$-135$^\circ$ and bond
lengthes within the range from 1.3 {\AA} to 1.7 {\AA} and found many
structurally intriguing C-sp$^2$-TDTCNs including all the previously
proposed cubic ones (some of them are selected to show in Fig2).
Especially, C-sp$^2$-TDTCNs can be found in every space group but
some of them degenerate to the ones with relatively higher symmetry.\\
\indent The summarized lattice constants and inequivalent atomic
positions for these C-sp$^2$-TDTCNs are based on the post-processes
of structure optimizations by regarding these C-sp$^2$-TDTCNs as
sp$^2$-hybridized carbon crystals. We have considered further energy
evaluation for these C-sp$^2$-TDTCNs. In the following, we introduce
our results of C-sp$^2$-TDTCNs templates for sp$^2$-like carbon
crystals, high pressure crystals of element nitrogen and TD
hydrocarbon crystals.\\
\begin{table*}
  \centering
\footnotesize
  \caption{The nominations, characters or previous names (Ch/Pn), lattice constant (LC:{\AA}), inequivalent positions (POS),
  and cohesive energy (Ecoh: eV/atom) of C-sp$^2$-TDTCNs with two inequivalent atomic positions}\label{tabII}
\begin{tabular}{c c c c c c c c c}
\hline \hline
C-sp$^2$-TDTCNs  &Ch/Pn  &LC          &POS1           &POS2   &Ecoh     \\
\hline
230-96h-48g &New           &10.341&(-0.144,-0.326, 0.776)&(-0.301,-0.875, 0.551)&-6.824\\
229-96l-96l &New           &12.534&(-0.389, 0.231,-0.311)&( 0.054,-0.222,-0.629)&-7.278\\
227-192i-192i&D96-D8       &18.048&( 0.339,-0.716,-0.606)&( 0.811,-0.412,-0.534)&-7.443\\
227-96g-32e  &D32-D8        &11.203&( 0.834,-0.166, 0.506)&( 0.387, 0.613, 0.613)&-7.191\cite{a}\\
227-192i-96g &D72-D9       &15.605&( 0.130, 0.431,-0.749)&(-0.281, 0.925,-0.781)&-7.253\\
226-192j-64g&New           &13.227&( 0.399,-0.447,-0.796)&( 0.621,-0.621,-0.379)&-7.191\\
225-48h-32f &C20\cite{c201, c202} &9.146 &(-0.139,-0.139, 0.139)&( 0.000,-0.198, 0.198)&-6.878\\
225-48g-32f &New           &9.572 &( 0.250, 0.430,-0.250)&( 0.850, 0.650,-0.650)&-6.819\\
225-192l-96k&New           &15.146&( 0.596,-0.839,-0.728)&( 0.455,-0.307,-0.307)&-7.219\\
225-192l-96j &New          &14.789&(-0.283, 0.352, 0.418)&(-0.224, 0.500, 0.341)&-7.289\\
224-48l-48l &D96-D8        &9.168 &( 0.728, 0.837, 0.947)&( 0.323, 0.572, 0.876)&-7.316\\
224-12g-8e  &New           &5.342 &( 0.000, 0.000, 0.362)&( 0.358, 0.642, 0.642)&-6.721\cite{a}\\
221-24m-24l &P48-P9        &8.197 &( 0.083, 0.353, 0.353)&( 0.500, 0.838, 0.278)&-7.124\\
221-48n-24l &P72-P8        &9.464 &( 0.840, 0.371, 0.734)&( 0.072, 0.500, 0.726)&-7.525\\
221-48n-24k &P72-P8        &9.462 &( 0.339, 0.234, 0.129)&( 0.000, 0.572, 0.774)&-7.521\\
220-24d-16c &New           &6.808 &( 0.234,-0.500, 0.250)&(-0.412,-0.412, 0.588)&-7.118\\
217-12e-8c  &Cage-N10\cite{n6}&4.979 &( 0.648, 0.000, 0.000)&( 0.324,-0.324,-0.324)&-6.249\\
215-6g-4e   &New           &4.904 &( 0.500, 0.500, 0.140)&( 0.693, 0.307, 0.307)&-6.365\\
215-24j-24j &New           &8.723 &( 0.076, 0.458, 0.347)&( 0.677, 0.789, 0.561)&-7.188\\
214-48i-48i &G48-G8        &10.532&( 0.347,-0.640, 0.196)&( 0.007,-0.669, 0.492)&-7.039\\
212-24e-8c  &New           &7.705 &( 0.803, 0.473, 0.305)&( 0.715, 0.715, 0.715)&-6.972\cite{a}\\
205-24d-8c  &New           &6.391 &( 0.410, 0.390, 0.672)&( 0.258, 0.242, 0.758)&-6.967\\
204-48h-24g &G36-G7        &9.365 &(-0.133,-0.342, 0.274)&( 0.285, 0.000, 0.572)&-7.517\\
200-12j-8f  &New           &5.451 &( 0.241, 0.369, 0.000)&( 0.223, 0.777, 0.777)&-7.067\\
195-6g-4e  &New           &4.1303 &( 0.000, 0.500, 0.322)&( 0.241, 0.759, 0.759)&-6.366\\
\hline \hline
\end{tabular}
\end{table*}
\subsection{C-sp$^2$-TDTCNs with one inequivalent position}
In our searching process, only seven C-sp$^2$-TDTCNs containing one
inequivalent atomic position were found. Although our searching
process is non-global, we still believe that there are only seven
C-sp$^2$-TDTCNs with one inequivalent atomic position under the
given requirements of cutoff radius, bond lengthes and bond angles.
We identify that six of them are the previously proposed 6.8$^2$G
(No.230)\cite{s2, p2}, interpenetrated-K4 (No.230, we refer it as
IK4)\cite{4, 8}, 6.8$^2$P (No.229) \cite{p1}, sp$^2$-diamond
(No.227)\cite{pccp}, 6.8$^2$D (No.224)\cite{p1} and K4
(No.214)\cite{k41, k42, k43} crystals. Their crystalline views
(fully optimized carbon crystals with first-principles method with
calculation details provided in the supplementary information) are
shown in FigS1. (a), (b), (c), (d), (e) and (f), respectively, in
the supplementary information. We name these C-sp$^2$-TDTCNs as
230-96h, 230-16b, 229-48k, 227-96g, 224-24i and 214-8a,
respectively, according to their space group numbers and Wyckoff
positions. Their corresponding lattice constants and inequivalent
atomic positions are summarized in Table SI in the supplementary
information. The seventh one belonging to 206 space group has not
been reported as a carbon crystal before but was proposed as the
3/6/c5 configuration by E. Koch et. al in 1995\cite{6} (The
structure of sp$^2$-diamond\cite{pccp} was also proposed as 3/6/c3
by E. Koch 1995\cite{6}). Its perspective views of both crystalline
and primitive cells are shown in FigS1 (g) and (h), respectively.
From its crystalline view, we can see that it contains eight
interconnected benzene rings in its cubic cell similar to that of
229-48k but in different connecting manner. We name it as 206-48e
according to its space group number and the fact that it contains
one inequivalent atomic position at the Wyckoff position of 48e (0.888, 0.985, 0.155).\\
\indent We have also found, in some other space group, many
C-sp$^2$-TDTCNs with only one inequivalent atomic position, which
degenerate to the seven ones with relatively higher symmetry. For
example, we found four equivalents for 224-24i (6.8$^2$D) in space
groups 228 (228-192h), 227 (227-192i), 215 (215-24j) and 201
(201-24h), respectively. More equivalent relations among these
C-sp$^2$-TDTCNs with only one inequivalent atomic position are shown
in Table SI. Furthermore, these seven inequivalent C-sp$^2$-TDTCNs
can also be found in the following two conditions of considering
about two and three inequivalent atomic positions in a cubic cell.
See the details of equivalent relations in all the C-sp$^2$-TDTCNs
in Table SII and SIII in the supplementary information.\\
\begin{table*}
  \centering
\footnotesize
  \caption{Nominations, characters or previous names (Ch/Pn), Lattice constant (LC:{\AA}), Inequivalent positions (POS),
  Cohesive energy (Ecoh: eV/atom) of C-sp$^2$-TDTCNs with three inequivalent atomic positions}\label{tabI}
\begin{tabular}{c c c c c c c c c c}
\hline \hline
C-sp$^2$-TDTCNs &Ch/Pn  &LC          &POS1           &POS2   &POS2 &Ecoh     \\
\hline
230-96h-96h-96h &G144-G8 &15.944 &(-0.054,-0.491, 0.569) &(-0.152,-0.559, 0.661) &(-0.099,-0.488, 0.644) &-7.531\\
230-48g-48f-16b &New    &9.882  &(-0.021,-0.375, 0.271) &( 0.000,-0.825, 0.451) &(-0.125,-0.375, 0.375) &-7.034\\
230-96h-96h-48f &New    &12.038 &(-0.035,-0.342, 0.501) &(-0.134,-0.400, 0.531) &( 0.000,-0.250, 0.564) &-7.283\\
229-96l-48k-16f &New    &11.844 &( 0.229, 0.638, 0.059) &( 0.413, 0.263,-0.263) &( 0.317, 0.317, 0.317) &-7.125\\
229-48j-48k-48k &New    &11.739 &( 0.241,-0.156, 0.000) &( 0.327,-0.187,-0.187) &( 0.281,-0.111,-0.111) &-7.013\\
229-96l-48k-48k &P192-P8\cite{s3, s5}&14.901 &( 0.278,-0.047, 0.396) &( 0.289,-0.173, 0.289) &( 0.324,-0.091, 0.324) &-7.627\\
228-192h-96g-32b&New   &14.433 &(-0.326, 0.857, 0.542) &(-0.375, 0.698, 0.553) &(-0.375, 0.625, 0.625) &-7.175\\
227-96g-96g-32e &New   &12.652 &( 0.169,-0.533, 0.830) &( 0.068,-0.288, 0.788) &( 0.171,-0.171, 0.671) &-7.207\\
227-96g-96g-96g &FCC-(C$_{36}$)$_2$\cite{s9}   &17.491 &(-0.105, 0.605, 0.801) &(-0.244, 0.858, 0.858) &(-0.123, 0.778, 0.778) &-7.125\\
227-192i-192i-96g&D120-D7   &18.994 &(-0.402,-0.711,-0.005) &(-0.910,-0.804, 0.357) &(-0.725,-0.275,-0.030) &-7.518\\
227-192i-96g-32e &FCC-(C$_{40}$)$_2$\cite{s9}   &15.881 &(-0.083,-0.021, 0.703) &( 0.063,-0.326, 0.937) &( 0.108,-0.392, 0.892) &-7.324\\
227-192i-192i-192i&D144-D8   &23.623 &(-0.315, 0.273,-0.358) &(-0.225, 0.317,-0.397) &(-0.416, 0.209,-0.373) &-7.423\\
226-96i-64g-48e &New   &14.601 &( 0.500,-0.638,-0.421) &( 0.583,-0.583,-0.583) &( 0.500,-0.298,-0.500) &-6.660\\
225-96k-96k-32f &New   &13.221 &( 0.319,-0.319, 0.449) &( 0.377,-0.623, 0.744) &( 0.337,-0.163, 0.337) &-7.076\\
225-192l-192l-192l &New   &18.868 &(-0.383, 0.331,-0.278) &(-0.431, 0.336,-0.215) &(-0.592, 0.464,-0.792) &-7.363\\
225-192l-192l-96j  &New   &20.015 &(-0.376, 0.276, 0.327) &(-0.434, 0.231,-0.317) &(-0.316, 0.267, 0.500) &-7.098\\
224-48l-48l-24l &D144-D8 &11.814 &( 0.631, 0.546, 0.715) &( 0.450, 0.634, 0.794) &( 0.833, 0.417, 0.745) &-7.422\\
224-24k-24k-24k &D72-D7  &8.757  &( 0.291, 0.711, 0.103) &( 0.254, 0.057, 0.057) &( 0.011, 0.215, 0.215) &-7.124\\
224-24h-24k-24k &D72-D8  &10.214 &( 0.000, 0.821, 0.500) &( 0.116, 0.116, 0.447) &( 0.810, 0.359, 0.810) &-7.545\\
224-48l-48l-24j &New     &10.245 &( 0.648, 0.855, 0.947) &( 0.053, 0.751, 0.849) &( 0.952, 0.548, 0.750) &-7.212\\
223-48l-24k-12h &New     &9.742  &( 0.308, 0.692, 0.869) &( 0.361, 0.634, 0.007) &( 0.431, 0.500, 0.000) &-7.115\\
221-24m-24l-8g  &P56-P8  &8.658  &( 0.639, 0.639, 0.161) &( 0.081, 0.500, 0.694) &( 0.696, 0.304, 0.304) &-7.170\\
221-48n-48n-24k &P120-P8 &11.971 &( 0.442, 0.279, 0.103) &( 0.370, 0.287, 0.203) &( 0.386, 0.301, 0.000) &-7.541\\
221-48n-48n-24l &P120-P8 &11.887 &( 0.704, 0.787, 0.871) &( 0.778, 0.603, 0.941) &( 0.885, 0.500, 0.799) &-7.499\\
217-48h-48h-48h &G72-G8  &11.982 &( 0.512,-0.132, 0.796) &( 0.094, 0.248, 0.603) &( 0.163, 0.244, 0.684) &-7.599\\
217-48h-48h-48h &6*A6$^3$12$^3$ &11.168 &( 0.491,-0.294,-0.141) &( 0.431,-0.342,-0.252) &( 0.411,-0.192,-0.099) &-7.215\\
217-48h-48h-24g &6A6   &10.504 &(-0.013,-0.179,-0.366) &(-0.099,-0.292,-0.387) &( 0.049,-0.347, 0.049) &-7.404\\
216-96e-48h-16e &New   &11.314 &( 0.293, 0.530,-0.618) &( 0.415, 0.585,-0.239) &( 0.858, 0.641,-0.641) &-7.228\\
216-96e-96e-48h &New   &13.869 &( 0.319,-0.608,-0.463) &( 0.251,-0.615,-0.545) &( 0.443,-0.285,-0.215) &-7.383\\
215-24j-12i-4e  &New   &7.589  &( 0.085, 0.448, 0.319) &( 0.533, 0.765, 0.765) &( 0.698, 0.302, 0.302) &-7.038\\
215-24j-24j-24j &6*A6$^3$12$^3$   &11.081 &( 0.636, 0.725, 0.545) &( 0.167, 0.466, 0.377) &( 0.054, 0.368, 0.534) &-6.996\\
215-24j-24j-24j &6*A6$^3$8$^3$   &9.231  &( 0.654, 0.763, 0.549) &( 0.069, 0.175, 0.459) &( 0.129, 0.429, 0.326) &-7.238\\
214-48i-48i-48i &G72-G8  &12.568 &(-0.121,-0.187, 0.721) &(-0.033,-0.629, 0.536) &(-0.029,-0.121, 0.696) &-7.263\\
212-12d-8c-4a   &New   &5.964  &( 0.295, 0.955, 0.125) &( 0.751, 0.751,-0.249) &( 0.125, 0.125, 0.125) &-7.055\\
212-24e-24e-8c  &New   &10.254 &( 0.813, 0.518, 0.313) &( 0.728, 0.563, 0.706) &( 0.702, 0.702, 0.702) &-6.485\\
207-24k-24k-8g  &P56-P7   &8.554  &( 0.958, 0.574, 0.317) &( 0.805, 0.426, 0.735) &( 0.706, 0.706, 0.706) &-7.299\\
206-48e-48e-48e &6*A7$^6$   &11.577 &( 0.156, 0.241,-0.669) &( 0.790, 0.125, 0.050) &( 0.901, 0.065, 0.030) &-7.466\\
206-48e-48e-48e &6*A6$^6$   &11.826 &( 0.085, 0.009, 0.814) &( 0.068, 0.128, 0.851) &( 0.156, 0.195, 0.795) &-7.347\\
206-48e-48e-48e &6*Z6$^6$   &10.526 &(-0.028,-0.402, 0.093) &(-0.101,-0.361, 0.214) &(-0.237,-0.296, 0.434) &-6.985\\
204-48h-48h-24g &New   &10.860 &(-0.162,-0.216, 0.344) &(-0.103,-0.568, 0.353) &( 0.561, 0.000, 0.395) &-7.062\\
203-96g-96g-32e &FCC-(C$_{28}$)$_2$\cite{s9} &15.028 &(0.393, 0.552, 0.823) &( 0.143, 0.451, 0.511) &( 0.575, 0.925, 0.575) &-7.095\\
201-24h-24h-8e  &D56-D7  &7.598  &( 0.787, 0.105, 0.649) &( 0.280, 0.903, 0.018) &( 0.147, 0.853, 0.147) &-7.044\\
200-12k-8i-6h   &New   &7.272  &( 0.500, 0.777, 0.345) &( 0.666, 0.666, 0.666) &( 0.500, 0.091, 0.500) &-6.979\\
200-24l-12j-8i  &New   &7.552  &( 0.165, 0.811, 0.591) &( 0.000, 0.090, 0.654) &( 0.725, 0.725, 0.725) &-6.821\\
200-12k-8i-6f   &New   &6.682  &(-0.305, 0.800, 0.500) &(-0.400, 1.000, 0.500) &(-0.317, 0.317, 0.317) &-6.952\\
200-12j-8i-6g   &New   &6.247  &( 0.279, 0.291, 0.000) &( 0.390, 0.500, 0.000) &( 0.195, 0.805, 0.805) &-6.642\\
197-24f-24f-8c  &P56-P7\cite{s2, p2}   &8.359  &(-0.098,-0.686, 0.297) &( 0.247,-0.419,-0.026) &( 0.236,-0.236, 0.236) &-7.183\\
195-12j-12j-6g  &New  &6.513  &( 0.952, 0.669, 0.271) &( 0.856, 0.618, 0.066) &( 0.000, 0.500, 0.389) &-6.845\\
\hline \hline
\end{tabular}
\end{table*}
\subsection{C-sp$^2$-TDTCNs with two inequivalent positions}
When we consider the situation of C-sp$^2$-TDTCNs containing two
inequivalent atomic positions, we have found 25 inequivalent
C-sp$^2$-TDTCNs. The perspective crystalline views of these 25
inequivalent C-sp$^2$-TDTCNs are shown in FigS2 in the supplementary
information and their corresponding lattice constants and
inequivalent atomic positions are summarized in Table I. Most of the
25 C-sp$^2$-TDTCNs are new except that the C20-carbon (225-48h-32f)
has been proposed as a carbon crystal\cite{c201, c202} and the
cage-N10 (217-12e-8c) has been proposed \cite{n6} as a high pressure
nitrogen phase. Some of them can be regarded as new Schwarzite
surfaces (indicated in Table I), such as 227-192i-192i (D surface
with 96 atomic positions in its primitive cell and octagon rings,
D96-D8), 221-24m-24l (P surface with 48 atomic positions in its
primitive cell and nonagon rings, P48-P9) and 204-48h-24g (G surface
with 36 atomic positions in its primitive cell and heptagon rings,
G36-G7). These C-sp$^2$-TDTCNs distribute in space groups of 230,
229, 227, 226, 225, 224, 221, 220, 217, 215, 214, 212, 205, 204,
200, and 195. Some of them can also be found in other space groups
or in the following condition of C-sp$^2$-TDTCNs containing three
atomic positions (see the equivalent relations in Table SII and SIII
in the supplementary information). In addition, we have also found
some C-sp$^2$-TDTCNs with two inequivalent atomic positions
degenerating to those with only one inequivalent atomic position.
Such as the one named as 228-96g-96g is equivalent to 224-24i
(6.8$^2$D). More examples can be found in Table SII in the supplementary information.\\
\subsection{C-TDTCNs with three inequivalent positions}
There are more C-sp$^2$-TDTCNs containing three inequivalent atomic
positions and the increase of the inequivalent atomic positions
enlarges the state space and reduces the probability of meeting the
special C-sp$^2$-TDTCNs in our searching process. Fortunately, we
have found 48 of them including some Schwarzite type surface
structures (indicated in Table II), such as the G type
230-96h-96h-96h (G144-G8) and P type 229-96l-48k-48k (P192-P8,
proposed before\cite{s3, s5}). Most of them are never mentioned in
literatures. Their corresponding lattice constants and inequivalent
positions are summarized in Table II and their corresponding
perspective crystalline views can be found in the supplementary Fig
S3 and Fig S4. Some of them are identified to be the previously
proposed FCC-(C$_{36}$)$_2$\cite{s9} (227-96g-96g-96g),
FCC-(C$_{40}$)$_2$\cite{s9} (227-192i-96g-32e),
FCC-(C$_{28}$)$_2$\cite{s9} (203-96g-96g-32e) and P56-P7\cite{s2,
p2} (195-12j-12j-6g). We notice that many new surface structures,
such as 227-192i-192i-96g (D120-D7), 227-192i-192i-192i (D144-D8),
224-48l-48l-24l (D144-D8), 221-24m-24l-8g (P56-P8), 221-48n-48n-24k
(P120-P8), 217-48h-48h-48h (G72-G8), and 201-24h-24h-8e (D56-D7)
have not been proposed previously. Some new structures belonging to
the same space and with the same atomic positions have the same name
(see 217-48h-48h-48h, 215-24j-24j-24j and 206-48e-48e-48e) and we
provide additional signs reflecting their structural characters for
distinguishing them. For example, 6*A7$^6$ means the structure with
characteristic hexagons connecting six heptagons at its six armchair
edges and 6*Z6$^6$ indicates characteristic hexagons connecting six
hexagons at its six zigzag vertexes.\\
\indent From table II, we can see that these C-sp$^2$-TDTCNs
distribute in space groups of 230, 229, 228, 227, 226, 225, 224,
223, 221, 217, 217, 216, 215, 214, 212, 207, 206, 204, 203, 201,
200, 197 and 195. This dose not means that C-sp$^2$-TDTCNs can not
be found in other space groups. For example, we found
227-96g-96g-32e in No.210 as 210-96h-96h-32e with relatively lower
symmetry. In addition, we have also found some C-sp$^2$-TDTCNs with
three inequivalent atomic positions degenerating to those with one
or two inequivalent atomic positions. Such as 213-8c-4a-4b to
230-16b (IK4) and 220-48e-48e-48e to
230-96h-48g. More examples can be found in Table SIII in the supplementary information.\\
\subsection{Additional discussions}
\indent All the post-processes of structure optimizations for these
C-sp$^2$-TDTCNs are performed with density functional theory (DFT)
based first-principles method by regarding these C-sp$^2$-TDTCNs as
sp$^2$-hybridized carbon crystals with calculation details provided
in the supplementary information. We should point out that, after
the optimization, a few C-sp$^2$-TDTCNs do not satisfy the
requirements of C-sp$^2$-TDTCNs in bond angles, bond lengths or
neighbours, which are understandable and acceptable. We have also
considered energy evaluation of these sp$^2$-hybridized carbon
crystals but have not performed systematic investigation on the
dynamical stability due to the large computational cost. Our results
show that some of them are energetically viable, such as
230-96h-96h-96h (G144-G8), 229-96l-48k-48k (P192-P8),
227-192i-192i-96g (D120-D7), 224-24h-24k-24k (D72-D8),
221-48n-48n-24k (P120-P8), 217-48h-48h-48h (G72-G8), 221-48n-24k
(P72-P8), 221-48n-24l (P72-P8), 204-48h-24g (G36-G7), and 224-24i
(6.8$^2$D) which possess remarkable energetic stability comparable
to that of diamond (-7.623 eV/atom). Most of them are more stable
than the experimentally produced graphdiyne (-7.06 eV/atom). In
addition, we want to point out that some of them may be dynamically
unstable like K4-carbon (214-16b) \cite{k43}, and some of them such
as sp$^2$-diamond and 6.8$^2$D have been confirmed
dynamically stable\cite{pccp}.\\
\section{Conclusion}
\indent Based on a simple restriction of lattice in cubic and
inequivalent atomic positions not exceeding three, we have
introduced a simple method for systematic searching of
C-sp$^2$-TDTCNs in all the cubic space groups from No.195 to No.230
and found many topological intriguing C-sp$^2$-TDTCNs. These
C-sp$^2$-TDTCNs are named according to their space group numbers and
Wyckoff positions, which can provide fundamental crystallographic
information for excluding future repeating structures. All of these
C-sp$^2$-TDTCNs can be considered as good templates for searching
for carbon crystals with novel properties, predicting high pressure
phases of element nitrogen and designing TD hydrocarbon crystals. We
believe that our results are of wide interests in both
mathematics and crystallography.\\
\section*{Acknowledgements}
\indent This work is supported by the National Natural Science
Foundation of China (Grant Nos. 51172191 and 11074211), the National
Basic Research Program of China (2012CB921303), the Program for New
Century Excellent Talents in University (Grant No.NCET-10-0169), and
the Scientific Research Fund of Hunan Provincial Education
Department (Grant Nos. 09K033, 10K065, 10A118).
\footnotesize{
\bibliography{rsc.bib} 

\begin{thebibliography}{35}
\bibitem{a} These structures, as carbon crystals, are three- and four-connected after optimization. We provide the final cohesive
energies and their initial configurations.
\end{thebibliography}
\bibliographystyle{}

\end{document}


\title{SUPPLEMENTARY INFORMATION for "Systematic
enumeration of crystalline networks with only sp$^{2}$ configuration
in cubic lattices"}
\author{Chaoyu He}
\affiliation{Hunan Key Laboratory for Micro-Nano Energy Materials
and Devices, Xiangtan University, Hunan 411105, P. R. China;}
\affiliation{Laboratory for Quantum Engineering and Micro-Nano
Energy Technology, Faculty of Materials and Optoelectronic Physics,
Xiangtan University, Hunan 411105, P. R. China.}
\author{L. Z. Sun}
\affiliation{Hunan Key Laboratory for Micro-Nano Energy Materials
and Devices, Xiangtan University, Hunan 411105, P. R. China;}
\affiliation{Laboratory for Quantum Engineering and Micro-Nano
Energy Technology, Faculty of Materials and Optoelectronic Physics,
Xiangtan University, Hunan 411105, P. R. China.}
\author{C. X. Zhang}
\affiliation{Hunan Key Laboratory for Micro-Nano Energy Materials
and Devices, Xiangtan University, Hunan 411105, P. R. China;}
\affiliation{Laboratory for Quantum Engineering and Micro-Nano
Energy Technology, Faculty of Materials and Optoelectronic Physics,
Xiangtan University, Hunan 411105, P. R. China.}
\author{J. X. Zhong}
\email{zhong.xtu@gmail.com} \affiliation{Hunan Key Laboratory for
Micro-Nano Energy Materials and Devices, Xiangtan University, Hunan
411105, P. R. China;} \affiliation{Laboratory for Quantum
Engineering and Micro-Nano Energy Technology, Faculty of Materials
and Optoelectronic Physics, Xiangtan University, Hunan 411105, P. R.
China.}
\date{\today}
\pacs{61.50.Ks, 61.66.Bi, 62.50. -p, 63.20. D-}

\begin{abstract}
We optimized these C-sp$^2$-TDTCNs through regarding them as carbon
crystals. Both the lattice constant and atomic positions of these
C-sp$^2$-TDTCNs were fully optimized until the residual forces on
each carbon atom to be less than 0.02 eV/{\AA}. All the calculations
of structure optimizations and energetic stability evaluations were
performed using the density functional theory based VASP code
\cite{vasp} with the projected augmented wave (PAW) potential
\cite{paw}. The exchange and correlation are approximated by general
gradient approximation (GGA) developed by Perdew et al. \cite{GGA}.
The wave functions for all systems are expanded by plane-wave
functions with cutoff energy of 500 eV. The Brillouin zone sample
meshes based on the Monkhorst-Pack scheme are set to be denser
enough (less than 0.25 1/{\AA}) to ensure the accuracy of our
calculations.\end{abstract} \maketitle
\begin{table*}
  \centering
\footnotesize
  \caption{The nominations, relations, lattice constant (LC:{\AA}), inequivalent positions (POS),
  and cohesive energy (Ecoh: eV/atom) of C-sp$^2$-TDTCNs with only one inequivalent atomic position}\label{tabSI}
\begin{tabular}{c c c c c c c c}
\hline \hline
C-TDTCNs  &Relation      &LC          &POS           &Ecoh     \\
\hline
230-96h   &6.8$^2$G       &9.587  &( 0.117, 0.548, 0.424) &-7.391\\
230-16b   &IK4-carbon     &4.426  &(-0.125,-0.375, 0.375) &-6.346\\
229-48k   &6.8$^2$P       &7.843  &( 0.689, 0.689, 0.913) &-7.364\\
227-96g   &sp$^2$-diamond &9.667  &(-0.049, 0.549,-0.726) &-7.179\\
224-24i   &6.8$^2$D       &6.095  &( 0.250, 0.087, 0.589) &-7.585\\
214-8a    &K4-carbon      &4.126  &(-0.125,-0.375, 0.375) &-6.529\\
206-48e   &3/6/c5         &7.505  &( 0.345,-0.111, 0.015) &-6.798\\
\hline
228-192h  &6.8$^2$D       &12.191 &( 0.207, 0.043, 0.875) &-7.585\\
227-192i  &6.8$^2$D       &12.191 &( 0.625, 0.793, 0.543) &-7.585\\
223-48l   &6.8$^2$P       &7.843  &( 0.311, 0.087, 0.689) &-7.364\\
222-48i   &6.8$^2$P       &7.843  &( 0.311, 0.087, 0.689) &-7.364\\
220-16c   &IK4-carbon     &4.426  &(-0.125,-0.375, 0.375) &-6.346\\
215-24j   &6.8$^2$D       &6.095  &( 0.250, 0.087, 0.589) &-7.585\\
213-8c    &K4-carbon      &4.126  &(-0.125,-0.375, 0.375) &-6.529\\
212-8c    &K4-carbon      &4.126  &( 0.375, 0.375, 0.375) &-6.529\\
211-48j   &6.8$^2$P       &7.843  &( 0.689, 0.689, 0.913) &-7.364\\
210-96h   &sp$^2$-diamond &9.667  &(-0.049, 0.549,-0.726) &-7.179\\
206-16c   &IK4-carbon     &4.426  &(-0.125,-0.375, 0.375) &-6.346\\
204-48h   &6.8$^2$P       &7.843  &( 0.689, 0.689, 0.913) &-7.364\\
203-96g   &sp$^2$-diamond &9.667  &(-0.049, 0.549,-0.726) &-7.179 \\
201-24h   &6.8$^2$D       &6.095  &( 0.250, 0.087, 0.589) &-7.585\\
199-8a    &K4-carbon      &4.126  &(-0.125,-0.375, 0.375) &-6.529\\
\hline \hline
\end{tabular}
\end{table*}
\begin{table*}
  \centering
\footnotesize
  \caption{The nominations, relations, lattice constant (LC:{\AA}), inequivalent positions (POS),
  and cohesive energy (Ecoh: eV/atom) of C-sp$^2$-TDTCNs with two inequivalent atomic positions}\label{tabSII}
\begin{tabular}{c c c c c c c c c}
\hline \hline
C-TDTCNs  &Relation  &LC          &POS1           &POS2   &Ecoh     \\
\hline
228-192h-192h &6.8$^2$P    &15.685&( 0.043, 0.155, 0.845)&( 0.043, 0.155, 0.655)&-7.364\\
228-96g-96g &6.8$^2$D      &12.191&( 0.875, 0.289, 0.539)&( 0.291, 0.375, 0.041)&-7.585\\
227-96h-96h &6.8$^2$D      &12.191&( 0.467,-0.375, 0.783)&( 0.462, 0.712, 0.375)&-7.585\\
226-192j-192j &6.8$^2$P    &15.685&( 0.155, 0.957, 0.845)&( 0.207, 0.905, 0.905)&-7.364\\
221-24m-24m &6.8$^2$P      &7.843 &( 0.311, 0.087, 0.689)&( 0.413, 0.189, 0.811)&-7.364\\
220-48e-48e &6.8$^2$G      &9.587 &(-0.798,-0.363,-0.674)&(-0.617,-0.547,-0.076)&-7.391\\
219-96h-96h &6.8$^2$D      &12.191&( 0.125, 0.043, 0.793)&( 0.207, 0.043, 0.875)&-7.585\\
219-48g-32e &215-6g-4e     &9.808 &( 0.250, 0.070, 0.750)&( 0.154, 0.154, 0.846)&-6.365\\
218-24i-24i &6.8$^2$P      &7.843 &( 0.087,-0.311, 0.311)&(-0.087,-0.311, 0.311)&-7.364\\
218-12f-8e  &217-12e-8c    &4.979 &( 0.648, 0.000, 0.000)&( 0.324,-0.324,-0.324)&-6.249\\
217-24g-24g &6.8$^2$P      &7.843 &( 0.087,-0.311, 0.311)&(-0.087,-0.311, 0.311)&-7.364\\
216-48h-48h &sp$^2$-diamond&9.667 &(-0.049, 0.549,-0.726)&(-0.201, 0.476,-0.299)&-7.179\\
216-96i-96i &6.8$^2$D      &12.191&( 0.043, 0.125, 0.293)&( 0.043, 0.125, 0.793)&-7.585\\
214-8a-8b   &IK4-carbn     &4.426 &(-0.125,-0.375, 0.375)&( 0.125,-0.625, 0.625)&-6.346\\
214-48i-48i &6.8$^2$G      &9.598 &( 0.452, 0.424, 0.617)&( 0.576, 0.383, 0.548)&-7.391\\
213-4a-4b   &K4-carbon     &4.126 &( 0.625, 0.625, 0.625)&( 0.125,-0.625, 0.625)&-6.529\\
212-4a-4b   &K4-carbon     &4.126 &( 0.875, 0.625, 0.375)&( 0.625, 0.625, 0.625)&-6.529\\
210-96h-32e &227-96h-32e   &11.203&( 0.834,-0.166, 0.506)&( 0.387, 0.613, 0.613)&-7.191\\
209-48h-32f &C20           &9.146 &(-0.139,-0.139, 0.139)&( 0.000, -0.198,0.198)&-6.878\\
209-48i-32f &225-48g-32f   &9.572 &( 0.250, 0.430,-0.250)&( 0.850, 0.650,-0.650)&-6.819\\
208-24m-24m &6.8$^2$P      &7.843 &( 0.087,-0.311, 0.311)&( 0.587,-0.189, 0.189)&-7.364\\
208-12l-12l &6.8$^2$D      &6.095 &( 0.250, 0.913, 0.413)&( 0.250, 0.587, 0.087)&-7.585\\
207-24k-24k &6.8$^2$P      &7.843 &( 0.087,-0.311, 0.311)&( 0.587,-0.189, 0.189)&-7.364\\
205-8c-8c   &IK4-carbon    &4.426 &(-0.125,-0.375, 0.375)&(-0.125,-0.625, 0.125)&-6.346\\
205-24d-24d &206-48e       &7.505 &( 0.595, 0.235,-0.361)&( 0.095, 0.735, 0.139)&-6.798\\
203-96g-32e &227-96g-32e   &11.203&( 0.834,-0.166, 0.506)&( 0.387, 0.613, 0.613)&-7.191\\
202-48h-32f &C20           &9.146 &(-0.139,-0.139, 0.139)&( 0.000, -0.198,0.198)&-6.878\\
202-48g-32f &225-48g-32f   &9.572 &( 0.250, 0.430,-0.250)&( 0.850, 0.650,-0.650)&-6.819\\
201-24h-24h &6.8$^2$P      &7.843 &( 0.087,-0.311, 0.311)&( 0.587,-0.189, 0.189)&-7.364\\
201-12f-8e  &224-12g-8e    &5.342 &( 0.000, 0.000, 0.362)&( 0.358, 0.642, 0.642)&-6.721\\
198-4a-4b   &K4-carbon     &4.126 &( 0.875, 0.625, 0.375)&( 0.625, 0.625, 0.625)&-6.529\\
197-24g-24g &6.8$^2$P      &7.843 &( 0.087,-0.311, 0.311)&(-0.087,-0.311, 0.311)&-7.364\\
197-12e-8c  &Cage10-N      &4.979 &( 0.648, 0.000, 0.000)&( 0.324,-0.324,-0.324)&-6.249\\
196-48h-48h &sp$^2$-diamond&9.667 &(-0.049, 0.549,-0.726)&(-0.201, 0.476,-0.299)&-7.179\\
195-12j-12j &6.8$^2$D      &6.095 &( 0.250, 0.913, 0.413)&( 0.250, 0.587, 0.087)&-7.585\\
\hline \hline
\end{tabular}
\end{table*}
\begin{table*}
  \centering
\footnotesize
  \caption{Nominations, relations, Lattice constant (LC:{\AA}), Inequivalent positions (POS),
  Cohesive energy (Ecoh: eV/atom) of C-sp$^2$-TDTCNs with only one inequivalent atomic position}\label{tabSIII}
\begin{tabular}{c c c c c c c c c c}
\hline \hline
C-TDTCNs  &Relation  &LC          &POS1           &POS2   &POS3 &Ecoh     \\
\hline
228-192h-192h-192h &224-24h-24k-24k  &20.233 &(-0.343, 0.845, 0.577) &( 0.440, 0.784, 0.940) &( 0.501, 0.750, 0.910) &-7.545\\
226-192j-192j-192j &221-48n-24l      &18.927 &( 0.420, 0.315, 0.867) &( 0.464, 0.250, 0.863) &( 0.367, 0.314, 0.920) &-7.525\\
220-48e-48e-48e &230-96h-48g         &10.340 &(-0.144,-0.326, 0.776) &(-0.301,-0.875, 0.551) &( 0.394,-0.526, 0.576) &-6.824\\
216-48h-16e-16e &C20                 &9.125  &( 0.547, 0.453, 0.272) &( 0.373, 0.127, 0.627) &( 0.610, 0.390, 0.390) &-6.878\\
213-8c-4a-4b    &IK4-carbon          &4.426  &(-0.125,-0.375, 0.375) &( 0.375, 0.375, 0.375) &( 0.125,-0.625, 0.625) &-6.346\\
212-8c-4a-4b    &IK4-carbon          &4.426  &( 0.125,-0.625, 0.625) &(-0.125,-0.375, 0.375) &(-0.125,-0.625, 0.125) &-6.346\\
210-96h-96h-32e &227-96g-96g-32e     &12.652 &( 0.169,-0.533, 0.830) &( 0.068,-0.288, 0.788) &( 0.171,-0.171, 0.671) &-7.207\\
210-96h-96h-96h &227-96g-96g-96g     &17.491 &(-0.105, 0.605, 0.801) &(-0.244, 0.858, 0.858) &(-0.123, 0.778, 0.778) &-7.125\\
209-96j-96j-96j &225-192l-96k        &15.146 &( 0.596,-0.839,-0.728) &( 0.455,-0.307,-0.307) &( 0.404,-0.161,-0.272) &-7.219\\
209-96j-96j-32f &225-96k-96k-32f     &13.221 &( 0.319,-0.319, 0.449) &( 0.377,-0.623, 0.744) &( 0.337,-0.163, 0.337) &-7.076\\
207-24k-24k-24k &221-48n-24k          &9.462 &( 0.340, 0.234, 0.129) &( 0.000, 0.572, 0.774) &( 0.660, 0.766, 0.871) &-7.521\\
205-24k-24k-24k &221-48n-24k          &9.462 &( 0.340, 0.234, 0.129) &( 0.000, 0.572, 0.774) &( 0.660, 0.766, 0.871) &-7.521\\
203-96g-96g-32e &227-96g-96g-32e     &12.652 &( 0.169, 0.967, 0.330) &( 0.213, 0.713, 0.932) &( 0.171, 0.829, 0.671) &-7.207\\
203-96g-96g-96g &227-96g-96g-96g     &17.491 &(-0.105, 0.605, 0.801) &(-0.244, 0.858, 0.858) &(-0.123, 0.778, 0.778) &-7.125\\
202-96j-96j-96j &225-192l-96k        &15.146 &( 0.596,-0.839,-0.728) &( 0.455,-0.307,-0.307) &( 0.339,-0.096,-0.272) &-7.219\\
202-96i-96i-32f &225-96k-96k-32f     &13.221 &( 0.319,-0.319, 0.449) &( 0.377,-0.623, 0.744) &( 0.337,-0.163, 0.337) &-7.076\\
202-96i-32f-32f &200-12j-8f          &10.901 &( 0.250, 0.864, 0.932) &( 0.364, 0.864, 0.864) &( 0.135, 0.864, 0.864) &-7.067\\
201-24h-24h-24h &204-48h-24g         &9.365  &( 0.500, 0.928, 0.785) &( 0.726, 0.867, 0.658) &( 0.367, 0.842, 0.774) &-7.517\\
197-24f-24f-24f &204-48h-24g         &9.365  &( 0.500, 0.928, 0.785) &( 0.367, 0.842, 0.774) &( 0.342, 0.726, 0.867) &-7.517\\
195-12j-4e-4e  &200-12j-8f           &5.412  &( 0.499, 0.271,-0.891) &( 0.718, 0.718,-0.282) &( 0.285,-0.285,-0.715) &-7.067\\
\hline \hline
\end{tabular}
\end{table*}
\begin{figure*}
  \includegraphics[width=7in]{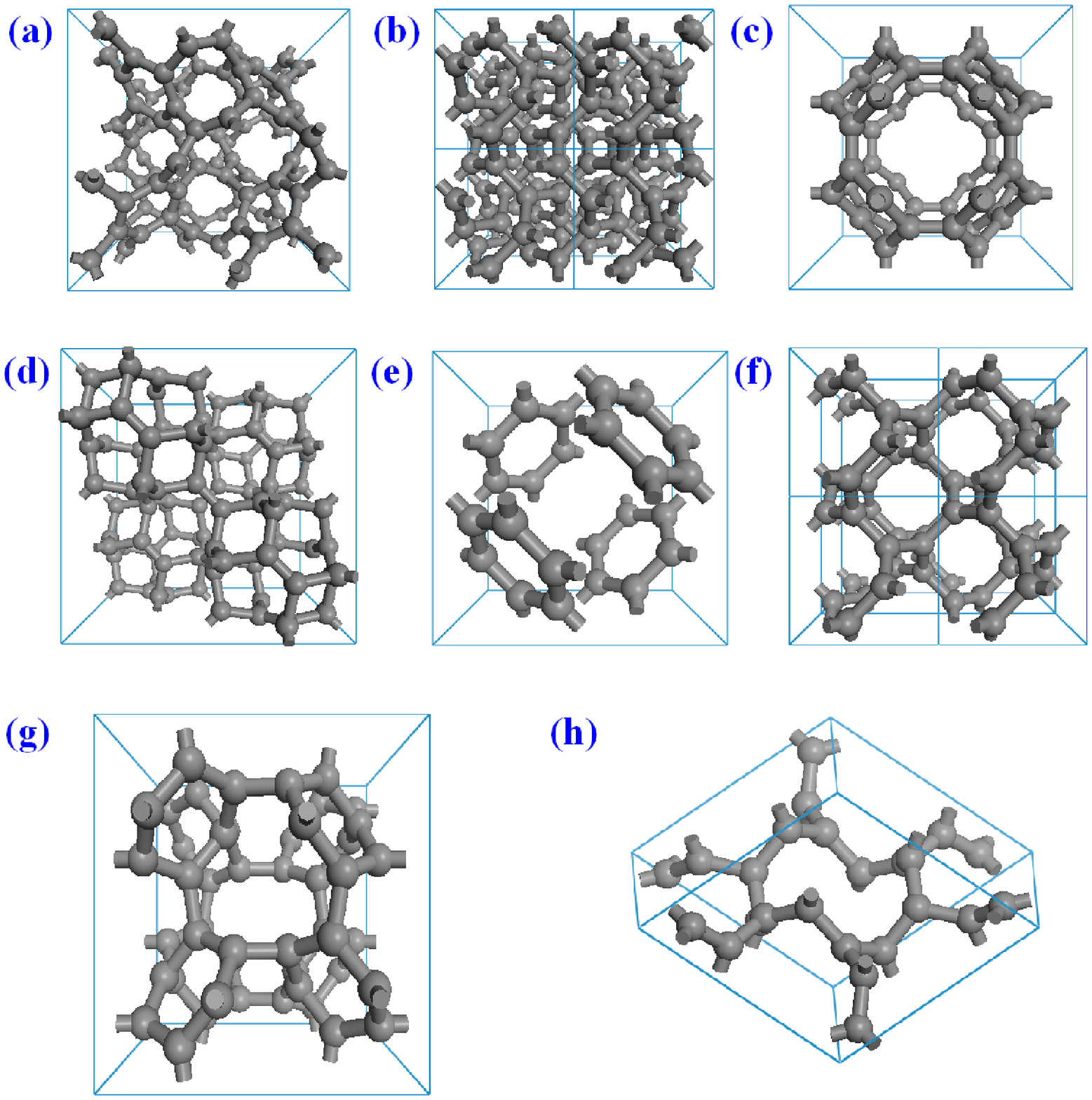}
  \caption{Crystalline views of the seven inequivalent C-sp$^2$-TDTCNs.
  (a), (b), (c), (d), (e) and (f) are 230-96h (6.8$^2$G), 230-16b (IK4), 229-48k (6.8$^2$P),
227-96g (sp$^2$-diamond), 224-24i (6.8$^2$D) and 214-8a (K4),
respectively. (g) and (h) are the crystalline views of 206-48e
(3/6/c5) in crystal cell and primitive cell,
respectively.}\label{FigS1}
\end{figure*}
\begin{figure*}
  \includegraphics[width=7in]{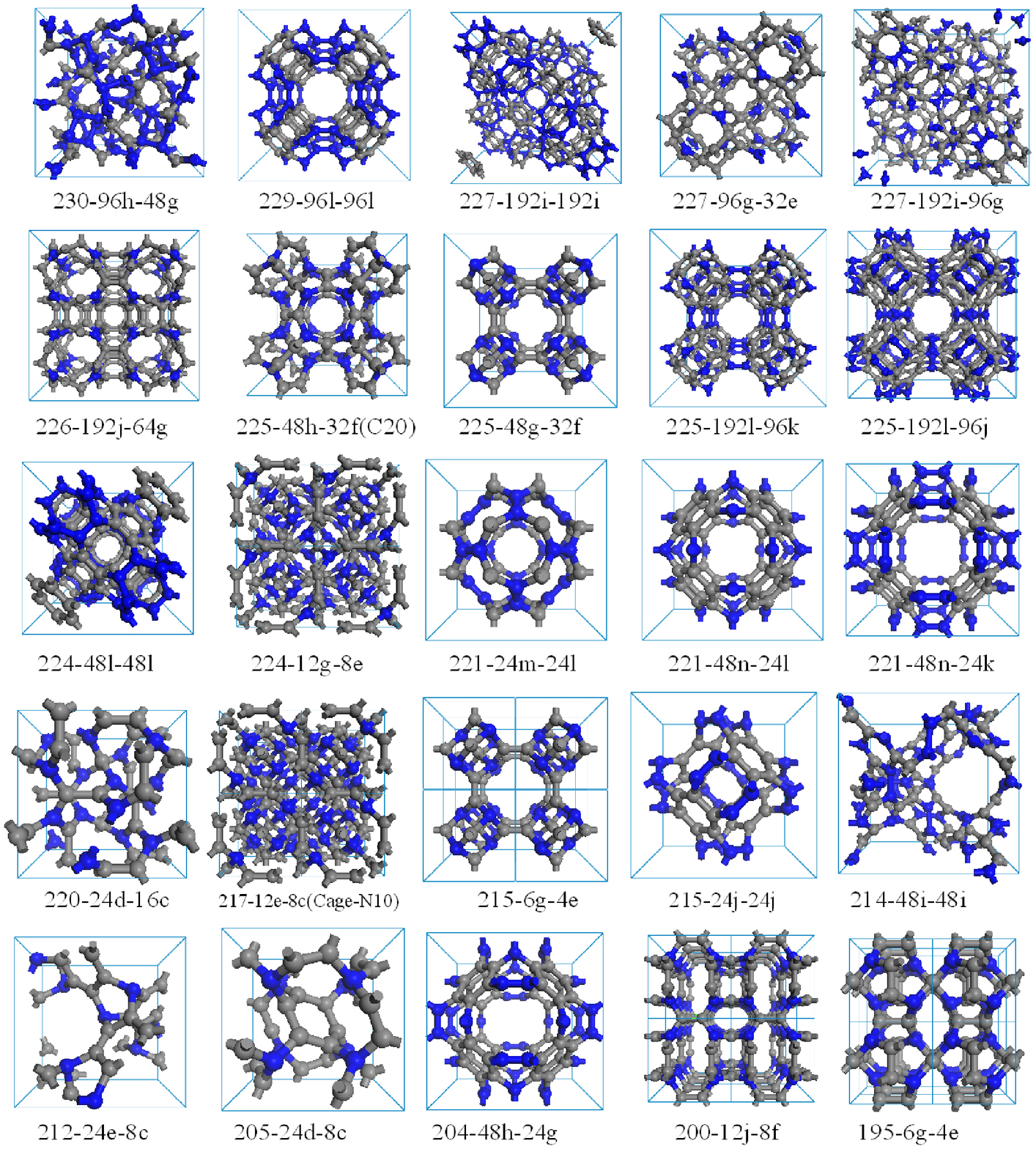}
  \caption{Crystalline views of the 25 inequivalent C-sp$^2$-TDTCNs with three inequivalent positions,
  where different atomic positions are shown in different colors}\label{FigS2}
\end{figure*}
\begin{figure*}
  \includegraphics[width=7in]{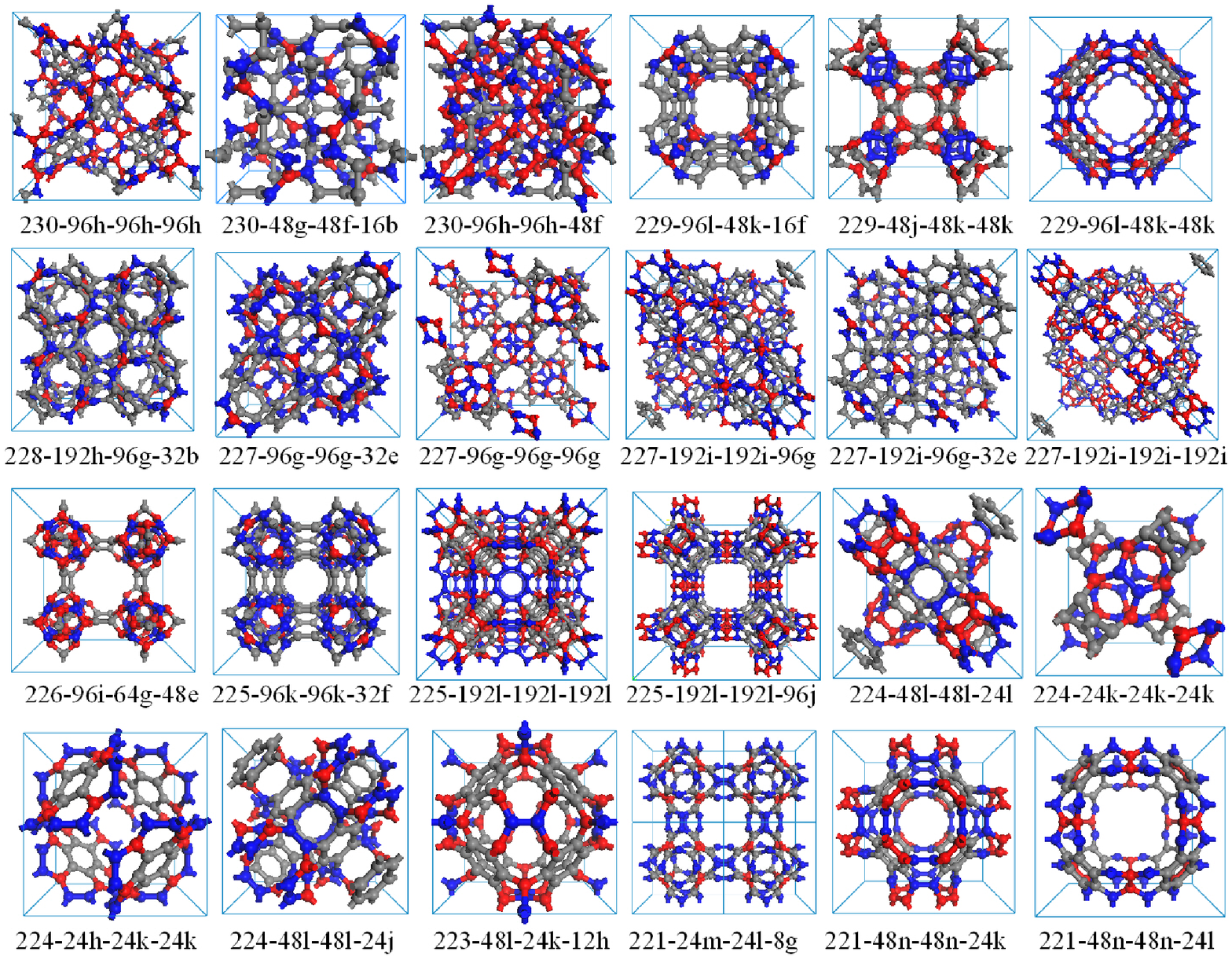}
  \caption{Crystalline views of the 24 inequivalent C-sp$^2$-TDTCNs (No.230-221) with three inequivalent positions,
  where different atomic positions are shown in different colors.}\label{FigS31}
\end{figure*}
\begin{figure*}
  \includegraphics[width=7in]{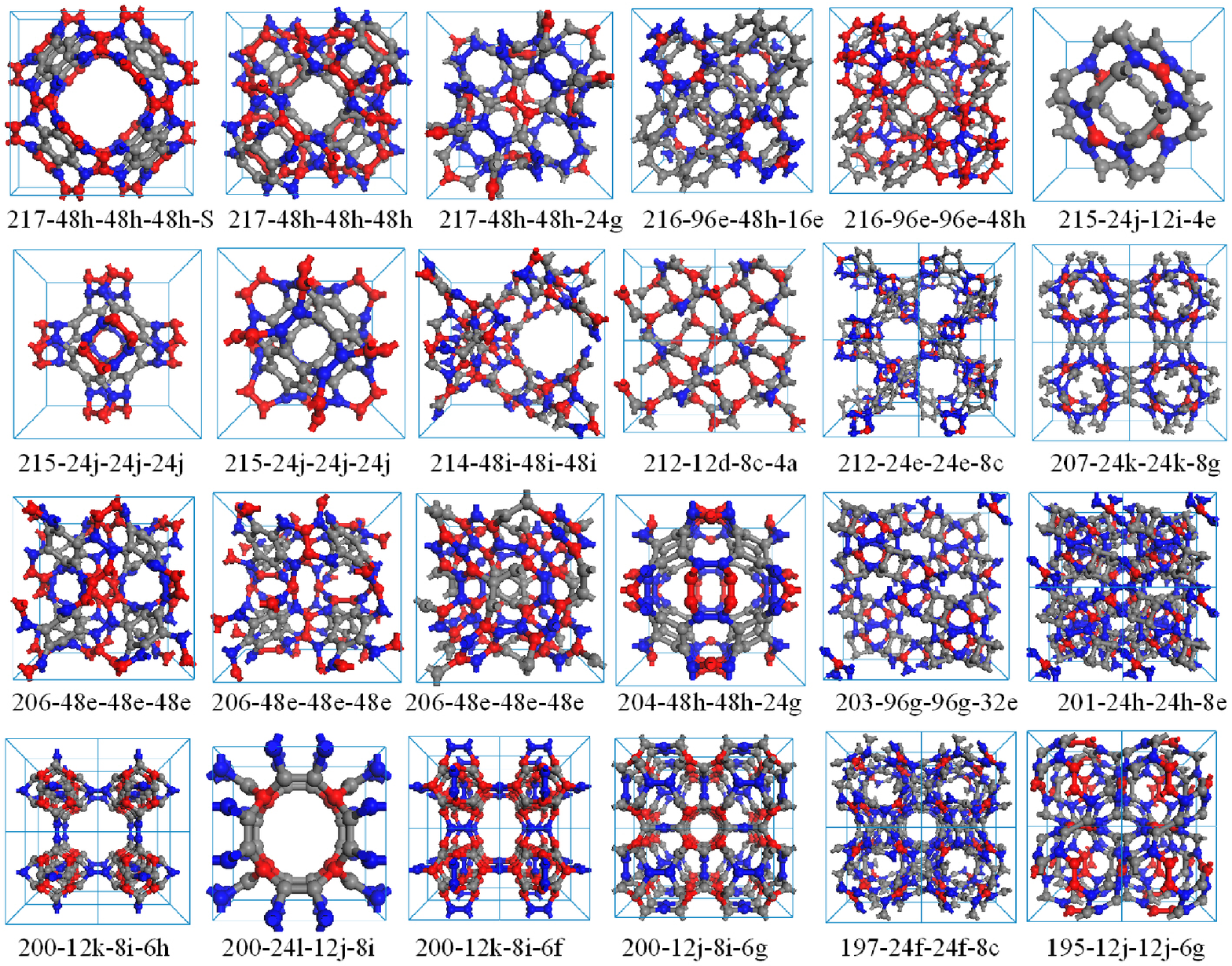}
  \caption{Crystalline views of the 24 inequivalent C-sp$^2$-TDTCNs (No.216-195) with three inequivalent positions,
  where different atomic positions are shown in different colors.}\label{FigS32}
\end{figure*}
\footnotesize{
\bibliography{rsc.bib} 
\bibliographystyle{}